\documentclass[aps, prd, amsmath, floats,floatfix, twocolumn,
superscriptaddress, nofootinbib, showpacs]{revtex4-1}

\usepackage{amssymb}
\usepackage{amsmath}
\usepackage{verbatim}
\usepackage{mathrsfs}
\usepackage{amsfonts}
\usepackage{latexsym}
\usepackage{epsfig}
\usepackage{color}
\usepackage{graphicx,subfigure}
\usepackage{units}
\usepackage{slashbox}
\usepackage{array}

\begin{document}

\title{Hints on halo evolution in SFDM models with galaxy observations}

\author{Alma X. Gonz\'alez-Morales} 
\affiliation{Instituto de Ciencias Nucleares, Universidad Nacional
  Aut\'onoma de M\'exico, Circuito Exterior C.U., A.P. 70-543,
  M\'exico D.F. 04510, M\'exico}

\author{Alberto Diez-Tejedor} 
\author{L. Arturo Ure\~na-L\'opez} 
\affiliation{Departamento de F\'isica, Divisi\'on de Ciencias e
  Ingenier\'ias, Campus Le\'on, Universidad de Guanajuato, Le\'on
  37150, M\'exico}

\author{Octavio Valenzuela} 
\affiliation{Instituto de Astronom\'ia, Universidad Nacional
  Aut\'onoma de M\'exico, Circuito Exterior C.U., A.P. 70-264,
  M\'exico D.F. 04510, M\'exico}


\date{\today}


\begin{abstract} 
A massive, self-interacting scalar field has been considered as a possible candidate for the dark matter in the universe. We present an observational
constraint to the model arising from strong lensing observations in galaxies. The result points to a discrepancy in the properties of scalar
field dark matter halos for dwarf and lens galaxies, mainly because halo parameters are directly related to physical quantities in the model.
This is an important indication that it becomes necessary to have a better understanding of halo evolution in scalar field dark matter models, where
the presence of baryons can play an important role.
\end{abstract}




\maketitle


\section{Introduction}
\label{sec:introduction}

The nature of dark matter (DM) remains elusive today, even though a generic cold particle weakly coupled to 
the standard model seems to be the most promising candidate~\cite{Bertone:2004pz,*Bergstrom:2012fi}. 
Treating DM as a bunch of classical particles is an appropriate effective description for many physical
situations. However, if DM is composed of bosons, the zero mode can develop a non-vanishing expectation 
value; this effect is usually known as Bose-Einstein condensation. A condensed phase does not admit a 
description in terms of classical particles, and the concept of a coherent excitation (i.e. a classical 
field) is more appropriate for practical purposes~\cite{Turner:1983he,*PhysRevLett.85.1158,*Matos:2000ss,*UrenaLopez:2008zh,*2012MNRAS.422..135R,*Chavanis:2011cz}. 
A specific realization of this scenario can be provided by the axion~\cite{Peccei:1977hh,*Sikivie:2009qn}, 
see also~\cite{Boehm:2003hm}.

In this paper we shall explore the lensing properties of a generic model of DM particles in a condensate, 
and compare the conditions necessary to produce strong lensing with those required to explain the dynamics 
of dwarf galaxies. As a result we will get some insight into halo evolution arising from this type of models.

In particular, we will consider the case of a complex, massive, self-interacting scalar field~$\phi$ 
satisfying the Klein-Gordon (KG) equation, $\Box\phi - (mc/\hbar)^2\phi -\lambda |\phi|^2\phi =0$, with the 
box denoting the d'Alembertian operator in four dimensions. For those natural situations in which the scalar
field mass $m$ is much smaller than the Planck scale, $m_{\textrm{Planck}}=(\hbar c/G)^{1/2}$, such that 
$\Lambda\equiv\lambda m_{\textrm{Planck}}^2/4\pi m^2 \gg 1$, the coherent (self-gravitating, spherically 
symmetric) solutions to the KG equation admit a very simple expression for the mass density~\cite{Colpi:1986ye,*Lee:1995af, Arbey:2003sj},
\begin{equation}
  \label{density}
  \rho(r)= \left\lbrace 
    \begin{array}{cl}
      \rho_c\, \displaystyle{\frac{\sin(\pi r/r_{\textrm{max}})}{(\pi
          r/r_{\textrm{max}})}} \quad & \textrm{for} \quad
      r<r_{\textrm{max}} \; \\
      0  \quad & \textrm{for} \quad r\ge r_{\textrm{max}} \; 
    \end{array}
  \right. \; .
\end{equation}
As usual we will refer to this model as scalar field dark matter (SFDM). Here $r_{\textrm{max}}\equiv \sqrt{\pi^2\Lambda/2}\,(\hbar/mc)$ 
is a constant with dimensions of length (notice that $r_{\textrm{max}}$ is just the Compton wavelength of 
the scalar particle, $\hbar/mc$, scaled by a factor of order $\Lambda^{1/2}$), and $\rho_c$ 
the density at the center of the configuration. The mass density profile in Eq.~(\ref{density}) leads to
compact objects of size $r_{\textrm{max}}$, and typical masses, $4\rho_c r_{\textrm{max}}^3/\pi$, that vary
from configuration to configuration according to the value of the central density. 

Eq.~(\ref{density}) was obtained without taking into account the gravitational influence of any other matter
sources, and assuming that all the scalar particles are in the condensate. It has been used as a first order 
approximation to describe the distribution of matter in dwarf shperoidals, which are expected to be DM 
dominated~\cite{Arbey:2003sj,Harko:2011xw,*Lora:2011yc,Robles:2012uy}. The mass distribution would be 
smooth close to the center of these galaxies, alleviating the cusp/core problem motivated by the discrepancies 
between the observed high resolution rotation curves and the profiles suggested by N-body
simulations~\cite{deBlok:2002tg}; see however \cite{Valenzuela:2005dh,*Adams:2011xq}.

The dynamics of dwarf galaxies suggests a self-interacting scalar with $m^4/\lambda \sim 50 - 75\,$(eV/c$^2$)$^4$, 
(i.e. $r_{\textrm{max}}\sim 5.5 - 7\,$Kpc), and typical central densities of the order of 
$\rho_c\sim 10^{-3}\, M_\odot/\textrm{pc}^{3}$, see Ref.~\cite{Arbey:2003sj}.
We are aware that Milky Way size galaxies are, at least, an order of magnitude larger than this value of 
$r_{\textrm{max}}$, and then they do not fit in this model as it stands. Nonetheless, if not all the DM 
particles are in the condensate, there is a possibility to have gravitational configurations where the inner
regions are still described by the mass density profile in Eq. (\ref{density}), wrapped in a cloud 
of non-condensed particles \cite{Bilic:2000ef,*Harko:2011dz}. For the purpose of this paper we do not need 
to specify the complete halo model. This is because strong lensing is not very sensitive to the mass 
distribution outside the Einstein radius, at most of the order of a few Kpc, just bellow the expected value 
of $r_{\textrm{max}}$. We could not neglect the exterior profile of the halo if we were interested, for instance, 
in weak lensing observations.

\section{Lensing properties of SFDM halos}\label{sec:lensing}

In the weak field limit the gravitational lensing produced by a mass distribution can be read directly from 
the density profile. As usual we assume spherical symmetry, and use the thin lens approximation, that is, 
the size of the object is negligible when compared to the other length scales in the configuration, i.e. the 
(angular) distances between the observer and the lens, $D_{\textrm{OL}}$, the lens and the source, $D_{\textrm{LS}}$, 
and from the observer to the source, $D_{\textrm{OS}}$.

Under these assumptions the lens equation takes the form
\begin{equation}
  \label{lens.eq}
  \beta = \theta -\frac{M(\theta)}{\pi D_{\textrm{OL}}^2 \theta
    \Sigma_{\textrm{cr}}} \; ,
\end{equation}
with $\beta$ and $\theta$ denoting the actual (unobservable) angular position of the source, and the apparent 
(observable) angular position of the image, respectively, both measured with respect to the line-of-sight~\cite{2002glml.book.....M}. 
The (projected) mass enclosed in a circle of radius $\xi$, $M(\xi)$, is defined from the (projected) surface 
mass density, $\Sigma (\xi)$, through
\begin{equation}
  \label{defs.sigma.m}
  \Sigma (\xi)\equiv \int_{-\infty}^{\infty}\! dz\, \rho (z,\xi)\, ,
  \quad M(\xi)\equiv 2\pi \int_0^\xi \! d\xi ' \, \xi ' \Sigma(\xi ') \, .
\end{equation}
Here $\xi=D_{\textrm{OL}}\theta$ is a radial coordinate in the lens plane, and $z$ a coordinate in the 
orthogonal direction. Finally $\Sigma_{\textrm{cr}}\equiv c^2 D_{\textrm{OS}} / 4\pi G D_{\textrm{OL}} D_{\textrm{LS}}$ 
is a critical value for the surface density. 

In general, Eq.~(\ref{lens.eq}) will be non-linear in $\theta$, and it could be possible that for a given 
position of the source, $\beta$, there would be multiple solutions (i.e. multiple images) for the angle 
$\theta$. This is what happens in the strong lensing regime to be discussed below. One particular case is 
that with a perfect alignment between the source and the lens, that actually defines the Einstein ring, with 
an angular radius of $\theta_{\textrm{E}}\equiv \theta(\beta=0)$. 

For a SFDM halo, and in terms of the normalized lengths $\xi_*\equiv\xi/r_{\textrm{max}}$ and  
$z_*\equiv z/r_{\textrm{max}}$, the surface mass density takes 
the form
\begin{equation}
  \label{sigma.sfdm}
  \Sigma_{\textrm{SFDM}}(\xi_*) =  \frac{2\rho_c r_{\textrm{max}}}{\pi}
  \int_{0}^{z_{\textrm{max}}}\,\frac{\sin (
    \pi\sqrt{\xi_*^2+z_*^2})}{\sqrt{\xi_*^2+z_*^2}}\, dz_* \; ,
\end{equation}
with $\quad 0\le \xi_* \le 1$ and $z_{\textrm{max}} = \sqrt{1 - \xi_*^2}$. A similar expression can be 
obtained for the mass enclosed in a circle or radius $\xi$, see Eq. (\ref{defs.sigma.m}) above. 
Here we are not considering the effect of a scalar cloud surrounding the condensate. For $r\lesssim r_{\textrm{max}}$ 
this will appear as a projection effect, which is usually considered to be small \cite{Kling:2007pw}. Indeed, 
we have corroborated that the inclusion of an outer isothermal sphere does not affect the conclusions 
of this paper.

With the use of the expression for the projected mass, $M_{\textrm{SFDM}}(\xi_*)$, the lens equation simplifies to
\begin{subequations}
  \label{lens.SFDM}
  \begin{equation}
    \beta_*(\theta_*)=\theta_* - \bar{\lambda}\,
    \frac{m(\theta_*)}{\theta_*} \; ,
\end{equation}
where $m(\xi_*)\equiv M_{\textrm{SFDM}}(\xi_*)/\rho_c r_{\textrm{max}}^3$ is a normalized mass function, 
evaluated numerically. Here $\beta_*=D_{\textrm{OL}} \beta/ r_{\textrm{max}}$ and $\theta_*=D_{\textrm{OL}}\theta/r_{\textrm{max}}$ 
are the normalized angular positions of the source and images, respectively, and the parameter 
$\bar{\lambda}$ is given by
\begin{equation}
    \label{lambda}
    \bar{\lambda} \equiv \frac{\rho_c r_{\textrm{max}}}{\pi
      \Sigma_{\textrm{cr}}} = 0.57 h^{-1}
    \left( \frac{\rho_c}{M_{\odot}\textrm{pc}^{-3}} \right) \left(
      \frac{r_{\textrm{max}}}{\textrm{kpc}} \right)
    \frac{d_{\textrm{OL}}d_{\textrm{LS}}}{d_{\textrm{OS}}} \; .
\end{equation}
\end{subequations}
In order to avoid confusion with the self-interaction term, $\lambda$, we have introduced a bar in the new 
parameter $\bar{\lambda}$. We have also defined the reduced angular distances $d_A\equiv D_{A} H_0 /c$,
and considered $H_{0}\equiv100h$(km/s)/Mpc as the Hubble constant today, with $h=0.710\pm 0.025$~\cite{Jarosik:2010iu}. 

\begin{figure}[t!]
\centering
\includegraphics[width=8cm]{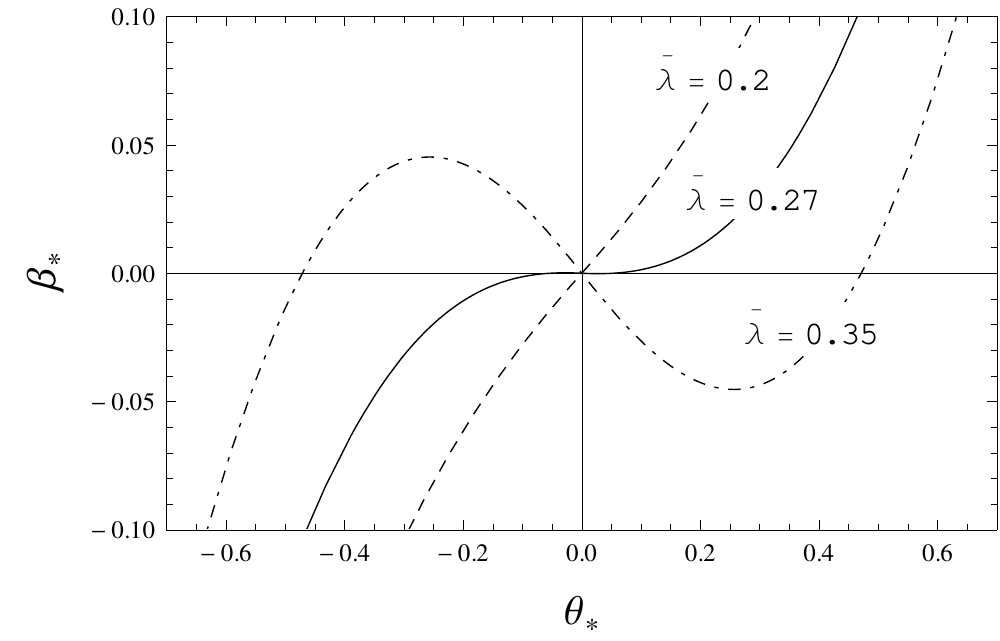}
\caption{The lens equation of a SFDM halo model, Eq. (\ref{lens.SFDM}), as a function of $\bar{\lambda}$.
The roots define the Einstein radius, $\theta_{*\textrm{E}}$, and its local maximum (minimum) the 
critical impact parameter, $\beta_{*\textrm{cr}}$. Both quantities are well
defined only for values of $\bar{\lambda} > \bar{\lambda}_{\textrm{cr}} \simeq 0.27$, which is the threshold value for strong lensing. 
}
\label{fig:SLSF}
\end{figure}

In Fig.~\ref{fig:SLSF} we show the behavior of the lens equation (\ref{lens.SFDM}) as the $\bar{\lambda}$ 
parameter varies (i.e. for different values of the combination $\rho_c r_{\textrm{max}}$). Some notes are in 
turn: $i)$~Strong lensing can be produced only for configurations with $\bar{\lambda} > \bar{\lambda}_{\textrm{cr}}
\simeq 0.27$, and $ii)$~For these configurations, only those with an impact parameter $|\beta_*|<\beta_{*\textrm{cr}}$ 
can produce three images (note that the actual value of $\beta_{*\textrm{cr}}$ depends on the parameter 
$\bar{\lambda}$, $\beta_{*\textrm{cr}}(\bar{\lambda})$).

These conditions on the SFDM profile are very similar to those obtained for the Burkert model in~\cite{Park:2003br}; 
this is not surprising because both of them have a core in radius. In that sense SFDM halos are analogous to 
those proposed by Burkert~\cite{1995ApJ...447L..25B}, but with the advantage that their properties are 
clearly connected to physical parameters in the model.

\begin{figure}[t!]
\centering
  \includegraphics[width=8cm]{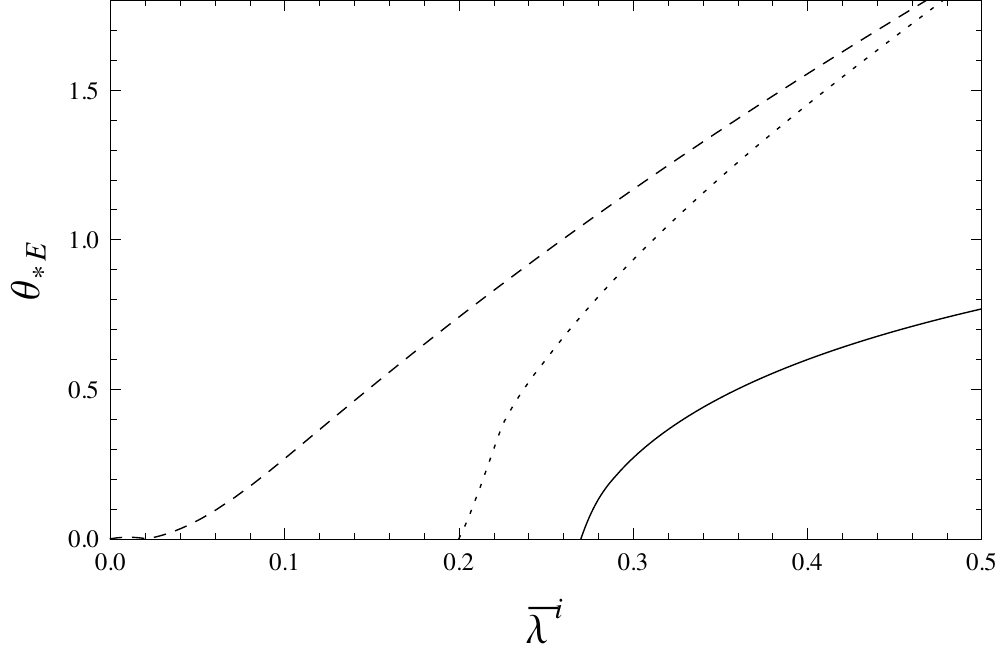} \\
\caption{The Einstein radius, $\theta_{*\textrm{E}}$, as function of $\bar{\lambda}^i$, for SFDM (solid line), NFW (dashed line), and Burkert (dotted
line) halo models. Einstein rings of similar magnitude require $\bar{\lambda}^{\textrm{NFW}} < \bar{\lambda}^{\textrm{Burkert}} <
\bar{\lambda}^{\textrm{SFDM}}$.}
\label{fig:combinedSF}
\end{figure}

In Fig.~\ref{fig:combinedSF} we show the magnitude of the Einstein radius, $\theta_{*\textrm{E}}$, as a 
function of the parameter $\bar{\lambda}$, where for comparison we have also plotted the same quantity for 
the NFW~\cite{2000ApJ...534...34W} and Burkert~\cite{Park:2003br} profiles. The minimum value of 
$\bar{\lambda}$ needed to produce multiple images is higher for a SFDM halo,
$\bar{\lambda}_{\textrm{cr}}^{\textrm{NFW}} = 0 <\bar{\lambda}_{\textrm{cr}}^{\textrm{Burkert}} = 2/\pi^2 <
\bar{\lambda}_{\textrm{cr}}^{\textrm{SFDM}} \simeq 0.27$. (Notice that there is an extra factor of $1/4\pi$ 
in our definition of $\bar{\lambda}$ when compared to that reported in Ref.~\cite{Park:2003br}.) SFDM halos 
seem to require larger values of $\bar{\lambda}$ in order to produce Einstein rings of similar magnitude to 
those obtained for the other profiles, but this is in part due to projection effects that have not been 
considered in this paper~\cite{Baltz:2007vq,Kling:2007pw}.

\section{Lensing VS Dynamics}\label{sec.constraints}

Taking into account that in SFDM models there is a critical value for the parameter $\bar{\lambda}$, 
$\bar{\lambda}_{\textrm{cr}}\simeq 0.27$, and considering the definition in Eq.~(\ref{lambda}), we can 
write the condition to produce strong lensing in the form
\begin{equation}
  \rho_c r_{\textrm{max}}[M_{\odot}\textrm{pc}^{-2}] \gtrsim 473.68 h
  f_{\textrm{dist}} \; , \label{constraint}
\end{equation}
with $f_{\textrm{dist}}\equiv d_{\textrm{OS}} / d_{\textrm{OL}}d_{\textrm{LS}}$ a distance factor.

In order to evaluate the right-hand-side (r.h.s.) of Eq.~(\ref{constraint}), we consider two surveys of 
multiply-imaged systems, the CASTLES~\cite{Falco:1999mb} and the SLACS~\cite{Bolton:2008xf}. From them we 
select only those elements for which the redshifts of the source and the lens have been determined (which 
amounts to approximately 60 elements in each survey), and calculate the corresponding distance factor~$f_{\textrm{dist}}$ 
for every element in the reduced sample. In CASTLES (SLACS) the distance factors are in the interval 
$4\lesssim f_{\textrm{dist}}\lesssim 27$, ($6\lesssim f_{\textrm{dist}}\lesssim 25$), with a mean value of 
$f_{\textrm{dist}}\simeq 7$, ($f_{\textrm{dist}}\simeq 11$), and then the r.h.s. of Eq.~(\ref{constraint}) 
takes on values in the range $1400 - 9000$, ($2000 - 8500$). Some representative elements from SLACS are 
shown in Table~\ref{tab:dynmic} (galaxy lensing). In terms of the mean values, the inequality in 
Eq.~(\ref{constraint}) translates into
\begin{subequations}\label{inequalities}
\begin{eqnarray}
 \rho_c\,r_{\textrm{max}}[M_{\odot}\textrm{pc}^{-2}]\gtrsim 2000\;
 ,\quad && \textrm{(CASTLES)} \\
 \rho_c\,r_{\textrm{max}}[M_{\odot}\textrm{pc}^{-2}]\gtrsim 4000\;
 ,\quad && \textrm{(SLACS)}
\end{eqnarray}
\end{subequations}

These numbers are an order of magnitude greater than those obtained from dwarf galaxies dynamics, 
$\rho_c r_{\textrm{max}}[M_{\odot}$pc$^{-2}] \simeq 100$, when interpreted using the same density 
profile~\cite{Harko:2011xw,*Lora:2011yc}; see again Table~\ref{tab:dynmic}. This is the main result of the paper. 
Remember that the value of $r_{\textrm{max}}$ is related to the fundamental parameters of the model, which 
are the mass of the scalar particle and the self-interaction term, and it remains constant throughout the 
formation of cosmic structure.

\begin{table*}[htp]
\begin{center}
    \begin{tabular}{ !{\vrule width 1 pt}c!{\vrule width 1 pt}c!{\vrule width 2pt}c!{\vrule width 1pt}r|c!{\vrule width 1pt}}
\noalign{\hrule height 1pt}
    \multicolumn{2}{ !{\vrule width 1 pt}c!{\vrule width 2pt}}{{\bf $\;\;$ DYNAMICS OF GALAXIES $\;\;$}} & \multicolumn{3}{c !{\vrule width 1 pt}}{{\bf
GALAXY LENSING}} \\ \noalign{\hrule height 1pt}
    Galaxy & $\;$ $\rho_c
r_{\textrm{max}}[M_{\odot}\textrm{pc}^{-2}]$ $\;$ & Galaxy & $f_{\textrm{dist}}$ $\;\,$ & $\;$ $\rho_c r_{\textrm{max}}[M_{\odot}\textrm{pc}^{-2}] $
$\;$\\  \noalign{\hrule height 1pt}
    Ho II & 	        $\;$36.19   & J0008$-$0004& 6.61 $\;$ & 2029.68 \\ \hline
    DD0 \,154 & 		$\;$66.47   & J1250+0523 & 8.46 $\;$ & 2832.41 \\ \hline
    DDO \,53 & 		$\;$67.53   &  J2341+0000 & 9.12 $\;$ & 3053.38 \\ \hline
    IC2574 & 	        $\;$81.89   & J1538+5817 & 11.74 $\;$ & 3930.44  \\ \hline
    NGC2366 & 		$\;$85.45   & J0216$-$0813 & 13.03 $\;$ & 4362.44  \\ \hline
    $\;$ Ursa Minor $\;$ & 104.72 &J1106+5228 & 15.74 $\;$ & 5269.75  \\ \hline
    Ho I & 		120.23  & J2321$-$0939 & 16.23 $\;$ & 5433.80 \\ \hline
    DD0\, 39 & 		145.94  & $\;$ J1420+6019 $\;$ & $\;$ 19.72 $\;$ & 6602.26 \\ \hline
    M81\,dwB &		265.87  & J0044+0113 & 25.26 $\;$ & 8457.05 \\ \noalign{\hrule height 1pt}
    \end{tabular}
\end{center}
\caption{
Estimates of the product $\rho_c r_{\textrm{max}}$ for different galaxies. {\it Left.}~As reported in Refs. \cite{Harko:2011xw,Lora:2011yc}, using
galactic dynamics. {\it Right.}~Derived from equation (\ref{constraint}) in this paper; recall that these values represent a lower limit (here we show
only a representative subsample of the SLACS survey). Note the difference of an order of magnitude between the values of $\rho_c r_{\textrm{max}}$ for dwarf galaxies
in the local universe, and the lower limit of this same quantity for galaxies producing strong lensing at 
$z\sim 0.5$. 
}
\label{tab:dynmic}
\end{table*}

We must recall that inequalities in Eq. (\ref{inequalities}) do not take into account the presence of baryons 
in galaxies. Gravity does not distinguish between luminous and dark matter; then the contribution of the 
former to the lens could be significant in some cases. For instance, for those systems in SLACS the stellar 
mass fraction within the Einstein radius is $0.4$, on average, with a scatter of $0.1$~\cite{Auger:2009hj}. 

We have corroborated that our estimates in  Eq. (\ref{inequalities}) are not sensitive to the inclusion of a 
baryonic component. To see that we add the contribution of a de Vacouler surface brightness 
profile~\cite{maoz:1993ix} to the lens equation, 
\begin{subequations}\label{lensing.new.com}
\begin{equation}\label{lensing.new}
    \beta_*(\theta_*)=\theta_* -\bar{\lambda}\,\frac{m(\theta_*)}{\theta_*} - \bar{\lambda}_{\textrm{lum}}\,
    \frac{f(\theta_{*}/r_{e*})}{\theta_*} \; .
  \end{equation}
Here $\bar{\lambda}_{\textrm{lum}}$ is a parameter analogous to that given in Eq~(\ref{lambda}), 
\begin{equation}
 \bar{\lambda}_{\textrm{lum}} \equiv \frac{(M/L)L}{2\pi \Sigma_{\textrm{cr}}} \; ,
\end{equation}
and $f(x)$ a dimensionless projected stellar mass,
\begin{eqnarray}
 f(x)=\frac{1}{2520} & & \left[e^q \left(q^7-7\,q^6 + 42\,q^5 - 210\,q^4 + 840\,q^3 \right.\right. \nonumber \\
 & & \left. \left.  -2520 q^2 + 5040 q -5040\right) + 5040\right] \; ,
\end{eqnarray}
\end{subequations}
with $q\equiv -7.76\, x^{-1/4}$. The mass-to-light ratio, $M/L$ (from a Chabrier initial mass function), and 
the effective radius, $r_e$, for each system in SLACS are reported in Ref.~\cite{Auger:2009hj}. 
With the use of Eq. (\ref{lensing.new.com}) strong lensing is always possible. Consequently, we must impose 
a different condition to constrain the product $\rho_c r_{\textrm{max}}$ in each galaxy, such as demand the 
formation of Einstein rings of certain radius.  

To proceed we use a small subsample of SLACS that includes configurations with the minimum, maximum, and 
mean Einstein radius, and stellar surface mass density, respectively. This is because the new lens equation 
is a function of the ratio $r_{e*} = r_e/r_{\textrm{max}}$; then, to compute the magnitudes of the Einstein 
radii, we shall fix the value of $r_{\textrm{max}}$ a priori. 

Using the new lens equation we find the value of $\bar{\lambda}$ that produces the appropriate Einstein radius 
for each of the elements in the subsample. This is done using two different values of 
$r_{\textrm{max}}$: $5$ and $10\,$Kpc. The resultant products $\rho_c r_{\textrm{max}}$ are compatible (in
order of magnitude) with the inequalities obtained from Eq. (\ref{constraint}). Only for those systems in 
the subsample with a high stellar surface mass density the value of $\rho_c r_{\textrm{max}}$ can decrease 
substantially, but it is important to have in mind that all the possible uncertainties associated to the 
distribution of the luminous matter, like the choice of the stellar initial mass function, will be more 
relevant in such cases. In general, these estimations are sensitive to the details of the particular 
configuration, and a more exhaustive  analysis, considering the complete sample, will be presented elsewhere.

\section{Discussion and final remarks} \label{sec:discussion}

We have shown that a discrepancy between lensing and dynamical studies appears if we consider that the SFDM 
mass density profile in Eq.~(\ref{density}) describes the inner regions of galactic halos at different
redshifts, up to radii of order $5-10\,$Kpc. More specifically, we have found that lens galaxies at 
$z\sim 0.5$, if correctly described by a SFDM halo profile, should be denser than dwarf spheroidals in 
the local universe, in order to satisfy the conditions necessary to produce strong lensing. 

In principle nothing guarantees that halos of different kind of galaxies share the same physical properties. 
Our studies took into account galaxies that are intrinsically different in terms of their total mass and 
baryon concentration. While dwarf galaxies show low stellar surface  brightness, stellar component in
massive, early type galaxies is typically compact and dense.

In the standard cosmological model the evolution of DM halos may trigger differences in concentrations for 
halos with different masses due to differences in the assembling epoch; smaller halos collapsed in an 
earlier and denser universe, therefore they are expected to be more concentrated. However, it is also well 
known that the presence of baryons during the assembly of galaxies can alter the density profile of the host 
halos and modify this tendency, making them shallower (supernova feedback~\cite{Ceverino:2007mi,*Governato:2009bg}), 
or even cuspier (adiabatic contraction~\cite{Blumenthal:1986ApJ}). Therefore, the stellar distribution may 
reveal different dynamical evolution for low and high mass halos triggered by galaxy formation. 

For SFDM, the dynamical interaction between baryons and the scalar field may also modify the internal halo 
structure predicted by the model, Eq.~(\ref{density}), clarifying the  discrepancy. For instance, if
the concentration of stellar distribution were correlated with that of the halo, like in the adiabatic 
contraction model when applied to standard DM halos, this may explain our findings. 
But at this time it is unknown how compressible SFDM halos are, and if such effect will be enough
to explain our results, because there are no predictions on its magnitude. 
If the modification triggered by 
baryons were insufficient,
then it might be suggesting an intrinsic evolution of SFDM halos across cosmic time. 
For example, if big galaxies emerge as the result of the collision of smaller ones, then the central 
densities of the resultant galaxies would be naturally higher; after all, $r_{\textrm{max}}$ is a constant 
in the model, and one would expect that total mass is preserved in galaxy-galaxy mergers. 
At this point we do not know which of these two mechanisms, 
the intrinsic to the model, or that due to the evolution of SFDM halos in the presence of baryons, is the dominant one. 
In that sense, a theoretical description of these processes may be very useful and welcome.

A full picture requires a distribution of values for the central density generated from the evolution of the 
spectrum of primordial density perturbations after inflation. Such a result is not available now, but it 
is possible to start tracing this distribution with galaxy observations. We present, 
for the first time, observational constraints on the dynamical evolution of SFDM halos in the presence of baryons, that must be considered 
for future semi-analytical/numerical studies of galaxy formation.




\begin{acknowledgments}
We are grateful to Juan Barranco for useful comments. This work was
partially supported by PROMEP, DAIP-UG, CAIP-UG, PIFI, 
I0101/131/07 C-234/07 of the Instituto Avanzado de
Cosmologia (IAC) collaboration, DGAPA-UNAM grant No. IN115311, 
and CONACyT M\'exico under grants 167335, 182445.
AXGM is very grateful to the members of the Departamento de F\'isica at Universidad de Guanajuato for their
hospitality.
\end{acknowledgments}


\bibliography{lensingrefs}


\end{document}